\documentclass[preprint,prd,amsmath,amssymb,floatfix,nofootinbib]{revtex4}

\usepackage[colorlinks=true,urlcolor=black]{hyperref}

\hypersetup{backref = true,pagebackref = true,hyperindex = true, colorlinks = true,
  breaklinks = true, urlcolor = blue, linkcolor = blue, bookmarks = true,
  bookmarksopen = true, citecolor=red}

\usepackage{bm}
\usepackage{color}
\usepackage{dcolumn}
\usepackage{graphics}
\usepackage{graphicx}
\usepackage{subfigure}
\usepackage{slashed}
\usepackage{placeins}

\usepackage{pdfpages}
\usepackage{eso-pic}
\usepackage{everyshi}
\usepackage{pdflscape}

\newcommand{\bea}{\begin{eqnarray}}
\newcommand{\eea}{\end{eqnarray}}
\newcommand{\be}{\begin{equation}}
\newcommand{\ee}{\end{equation}}

\newcommand{\ar}{a_s}

\begin{document}

\title{Gross-Llewellyn Smith sum rule with analytic coupling
}
\author{I.R. Gabdrakhmanov$^{1}$, N.A Gramotkov$^{1,2}$, A.V.~Kotikov$^{1}$,  O.V.~Teryaev$^{1}$, D.A. Volkova$^{1,3}$
  and I.A.~Zemlyakov$^{4,5}$}
\affiliation{
  $^1$Bogoliubov Laboratory of Theoretical Physics,
  Joint Institute for Nuclear Research, 141980 Dubna, Russia;\\
$^2$Moscow State University, 119991, Moscow, Russia;\\
  $^3$Dubna State University,
  141980 Dubna, Moscow Region, Russia;\\
  $^4$Department of Physics, Universidad Tecnica Federico Santa Maria,\\
  Avenida Espana 1680, Valparaiso, Chile,\\
$^5$Tomsk State University, 634010 Tomsk, Russia}


\begin{abstract}
  We investigate the Gross-Llewellyn Smith sum rule
  within the framework of analytic QCD.
A comparison is performed between experimental data, lattice calculations, and perturbative QCD
based on conventional and analytic versions of perturbation theory with different
parametrizations for the twist-four contribution. We show that conventional perturbation
theory
fails to reproduce the data, while
the analytic version
demonstrates good agreement with experiment
and lattice calculations. We also discuss the relation between the Gross-Llewellyn Smith sum rule and the Bjorken sum rule.

\end{abstract}

\maketitle

\section{Introduction}

The parity-odd structure function $F_3$, which contributes to deep inelastic neutrino scattering, plays an important role in understanding processes
related to both QCD
and electroweak physics. Its first Mellin moment determines the Gross-Llewellyn-Smith (GLS) sum rule \cite{Gross:1969jf}
\bea
C_{GLS}^{p+n}(Q^2)=\frac{1}{2}\, \int\limits^1_0\left[F_3^{\overline{\nu}p}(x,Q^2)+F_3^{\nu n}(x,Q^2)\right]dx\,,
 \label{GLS.I}
\end{eqnarray}
which in the quark-parton model counts the number of valence quarks in the proton. Perturbatively computed radiative corrections serve as an important
test of the suitability of QCD as a theory of strong interactions.

Since $F_3$ is a non-singlet quantity, the GLS sum rule has a vanishing anomalous dimension at leading order (LO) of perturbation theory.
The GLS sum rule is unique in that the nucleon matrix element of the leading twist
is known analytically,
and the Wilson coefficient is known up to next-to-next-to-next-to-leading order (N$^3$LO)
\cite{Retey:2000nq,Baikov:2010je}, making it,
along with the Bjorken sum rule (BSR) \cite{Bjorken:1966jh},
an ideal laboratory for determining the strong coupling constant $\alpha_s$
at the boundary of the nonperturbative regime.
However, the available experimental data, unlike the BSR case
(see the recent paper \cite{Deur:2021klh} and references therein),
are rather sparse. 
In light of this, 
recent lattice QCD calculations \cite{Can:2025zsr}
have indeed made valuable contributions to the study of the GLS sum rule.

The rest of the paper is organized as follows: in Sections II and III we present the basic formulas for the GLS and Bjorken sum rules and discuss
their similarities. 
We present and discuss our results in Section IV, and provide a summary and concluding remarks in Section V.

\section{Gross-Llewellyn Smith sum rule}

The GLS sum rule is defined by the integral (\ref{GLS.I}).
The quantity
$C_{GLS}^{p+n}(Q^2)$
can be expressed in the operator product expansion (OPE)
form:
\bea
C_{GLS}^{p+n}(Q^2)=
3 \, \bigl(1-D(Q^2)\bigr) +\sum_{n=2}\,\frac{\mu_{2n}}{Q^{2n-2}} \, ,
\label{Gpn.OPEab}
\eea
where $(1-D(Q^2))$ is the leading twist (or twist-two) contribution, and the coefficients $\mu_{2n}$
of the higher-twist terms are free parameters that must be determined from experimental data.

In this paper, we particularly focus on rather low $Q^2$ values, so
it is common to use 
the so-called "massive" twist-four representation (see \cite{Teryaev:2013qba,Gabdrakhmanov:2017dvg})
\footnote{
  Note that for relatively large values of $Q^2$ ($Q^2 \geq 1$ GeV$^2$), the "massive" twist-four term can be expanded
  in a series of inverse powers of $Q^2$, thereby restoring the usual form (\ref{Gpn.OPEab}) of the OPE.},
we have:
\bea
\overline{C}_{GLS}^{p+n}(Q^2)=
3 \, \bigl(1-D(Q^2)\bigr) +\frac{\hat{\mu}_4 M^2}{Q^{2}+M^2} \, .
\label{Gpn.mOPEa}
\eea

As in the previous case, the value 
of the twist-four term $\hat{\mu}_4$ is a free parameter that must be fitted from experimental data.

Up to the $k$-th order of perturbation theory (PT), the twist-two parts have the form
\bea
D^{(1)}(Q^2)=\frac{4}{\beta_0} \, a^{(1)}_s,~~D^{(k\geq2)}(Q^2)=\frac{4}{\beta_0} \, a^{(k)}_s\left(1+\sum_{m=1}^{k-1} d_m \bigl(a^{(k)}_s\bigr)^m
\right)
\label{GLS} 
\eea
where
\be
a^{(k)}_s(Q^2)=\frac{\beta_0 \alpha^{(k)}_s(Q^2)}{4\pi}=\beta_0\,\overline{a}^{(k)}_s(Q^2),
\label{ak}
\ee
and
$d_1$,
$d_2$,
and $d_3$
are exactly known (see \cite{Baikov:2010je} and references therein).
Hereafter, $\beta_0$ is the first coefficient of the QCD $\beta$-function:
\be
\beta(\overline{a}^{(k)}_s)\equiv  \frac{d \overline{a}^{(k)}_s(Q^2)}{dL}
=-{\left(\overline{a}^{(k)}_s\right)}^2 \bigl(\beta_0 + \sum_{i=1}^k \beta_i {\left(\overline{a}^{(k)}_s\right)}^i\bigr),
\label{bQCD}
\ee
where $\beta_i$ are known up to $k=4$ \cite{Baikov:2008jh}.


Following \cite{Cvetic:2006mk}, we introduce and utilize the derivatives (at the $k$-th order of PT)
\be
\tilde{a}^{(k)}_{n+1}(Q^2)=\frac{(-1)^n}{n!} \, \frac{d^n a^{(k)}_s(Q^2)}{(dL)^n},~~L=\ln \frac{Q^2}{\Lambda^2}\,,
\label{tan+1}
\ee
which play a key role in the construction of analytic QCD (see, e.g., \cite{Kotikov:2022swl}). 

The series of derivatives $\tilde{a}_{n}(Q^2)$ can be used instead of the series of $\ar$-powers. Indeed, although
each derivative reduces the $\ar$ power, it also introduces
an additional $\beta$-function and consequently an additional $\ar^2$ factor.
By definition (\ref{tan+1}), at LO, the expressions for $\tilde{a}_{n}$ and $\ar^{n}$ coincide exactly.
Beyond LO, there is a one-to-one correspondence between $\tilde{a}_{n}$ and $\ar^{n}$, established in \cite{Cvetic:2006mk,Cvetic:2010di} and
extended to the fractional case in \cite{GCAK}.


Converting the powers of the coupling constant
into its derivatives, we obtain
\bea
D^{(1)}(Q^2)=\frac{4}{\beta_0} \, \tilde{a}^{(1)}_s,~~D^{(k\geq2)}(Q^2)=\frac{4}{\beta_0} \, \tilde{a}^{(k)}_s\left(1+\sum_{m=1}^{k-1} \tilde{d}_m \bigl(\tilde{a}^{(k)}_s\bigr)^m
\right)
\label{GLS.1}
\eea
where
\be
\tilde{d}_1=d_1,~~\tilde{d}_2=d_2-b_1d_1,~~\tilde{d}_3=d_3-\frac{5}{2}b_1d_2-\bigl(b_2-\frac{5}{2}b^2_1\bigr)\,d_1,
\label{tdi} 
\ee
and $b_i=\beta_i/\beta_0^{i+1}$.

For the case of 3 active quark flavors ($f=3$), which is adopted in this work, the results are
\footnote{
  The coefficients $\beta_i$ $(i\geq 0)$ of the $\beta$ function (\ref{bQCD}) and hence the coupling constant $\alpha_s(Q^2)$ itself
  depend on the number $f$, and each new quark enters/leaves the game at a certain threshold $Q^2_f$ according to \cite{Chetyrkin:2005ia}.
  The corresponding parameters $\Lambda^{(f)}$ in N$^i$LO PT can be found in \cite{Chen:2021tjz}.}
\bea
d_1=\tilde{d}_1=1.59,~~d_2=3.75~(\tilde{d}_2=2.51),~~d_3=16.77~(\tilde{d}_3=10.44),
\, ,
\label{td123} 
\eea
i.e., the coefficients in the derivative series are slightly smaller.

{\it Heavy quark contribution}.
In addition to the light flavors, the GLS sum rule receives contributions from heavy quarks (HQs).

Taking the HQ effects into account (see \cite{Blumlein:2016xcy}), one obtains
\bea
&&C_{GLS}^{p+n}(Q^2)=
3 \, \left(1+\frac{|V_{cd}|}{1+\xi_c}-D(Q^2,\xi)\right) +\sum_{n=2}\frac{\mu_{2n}}{Q^{2n-2}} \, ,
\label{GLS.2abc} \\
&&\overline{C}_{GLS}^{p+n,(2)}(Q^2)=
3 \,  \left(1+\frac{|V_{cd}|}{1+\xi_c}-D(Q^2,\xi)\right)+\frac{\hat{\mu}_4 M^2}{Q^{2}+M^2} \, .
\label{GLS.2}
\eea
where
\be
\xi_i=\frac{Q^2}{m^2_i}~~(i=c,b,t)\,,
\label{xi} 
\ee
and
$m_i=m_i(Q^2=m_i^2)$ in the $\overline{MS}$-scheme, with
$m_c=1.27$ GeV, $m_b=4.18$ GeV, and $m_t=172.76$ GeV (see \cite{PDG20}).

Up to the $k$-th order of PT,
the twist-two parts have the form, similar to (\ref{GLS}),
\bea
&&D^{(1)}(Q^2,\xi)=\frac{4}{\beta_0} \, \tilde{a}^{(1)}_s\left(1-
\frac{|V_{cd}|}{6}\frac{2\ln(1+\xi_c)+1}{1+\xi_c}\right),~~\nonumber \\
&&D^{(k\geq2)}(Q^2,\xi)=
\frac{4}{\beta_0} \, \tilde{a}^{(k)}_s\left(1-
\frac{|V_{cd}|}{6}\frac{2\ln(1+\xi_c)+1}{1+\xi_c} + \tilde{d}_1(\xi) \tilde{a}^{(k)}_s
+\sum_{m=2}^{k-1} \tilde{d}_m \bigl(\tilde{a}^{(k)}_s\bigr)^m
\right),
\label{GLS.2aa}
\eea
where $\tilde{d}_1(\xi)$
is obtained from 
$d_1$ by the following redefinition:
\be
d_1 \to d_1 -  \frac{1}{3}\, \sum_{i=c,b,t}\,C_1(\xi_i),~~\tilde{d}_1 \to \tilde{d}_1 - \frac{1}{3}\,\sum_{i=c,b,t}\,C_1(\xi_i)\,,
    \label{Gpn.MA.HQ.GLS} 
    \ee
    where $C_1(\xi)$ has the following form:
\bea
C_1(\xi)&=&\frac{8}{3\beta_0} \, \biggl\{\frac{6\xi^2+2735\xi+11724}{5040\xi}- \frac{3\xi^3+106\xi^2+1054\xi+4812}{2520\xi}\,L(\xi)\nonumber\\ 
&-&\frac{5}{3\xi(\xi+4)}\,L^2(\xi)+\frac{3\xi^2+112\xi+1260}{5040}\,\ln(\xi)\biggl\}\,
\label{C1} 
\eea
and
\be
L(\xi)=\frac{1}{2\delta}\,\ln\left(\frac{1+\delta}{1-\delta}\right),~~\delta^2=\frac{\xi}{4+\xi}\,.
\label{L} 
\ee

The CKM-matrix element $|V_{cd}|=0.22486$ in (\ref{GLS.2aa}) is rather small.
Actually, there are also corresponding contributions
from $b$ and $t$-quarks, but they are multiplied by CKM-matrix coefficients
$|V_{bu}|^2$ and $|V_{td}|^2$, which have negligible values: $|V_{bu}|=0.00369$ and $|V_{td}|=0.00857$. Consequently,
such contributions $\sim |V_{bu}|$ and $\sim |V_{td}|$ in (\ref{GLS.2aa})
can be safely ignored.

Now, using the derivatives of the coupling constant introduced above (\ref{tan+1}), 
the results (\ref{Gpn.OPEab}) and (\ref{Gpn.mOPEa}) can be written in APT as:
\bea
&&C_{\rm{A},GLS}^{p+n}(Q^2)=
3 \, \bigl(1-D_{\rm A}(Q^2)\bigr) +\sum_{n=2} \frac{\mu_{2n,A}}{Q^{2n-2}} \, ,
\label{Gpn.MAab}\\
&&\overline{C}_{\rm{A},GLS}^{p+n}(Q^2)=
3 \, \bigl(1-D_{\rm A}(Q^2)\bigr) +\frac{\hat{\mu}_{4,A} M^2}{Q^{2}+M^2} \, ,
\label{Gpn.MAa}
\eea
where the perturbative part $D_{\rm{A}}(Q^2)$
takes the same forms as (\ref{GLS.1}),
but the derivatives $\tilde{a}^{(k)}_{\nu}$
  are replaced by the corresponding analytical derivatives $\tilde{A}^{(k)}_{\nu}$
  (the corresponding expressions for the analytic coupling constants
  $\tilde{A}^{(k)}_{\nu}$ can be found in \cite{Kotikov:2022sos,Kotikov:2023meh}):
\be
D^{(1)}_{\rm A}(Q^2)=\frac{4}{\beta_0} \, A^{(1)},~~
D^{k\geq2}_{\rm{A}}(Q^2) =\frac{4}{\beta_0} \, \Bigl(A^{(k)}
+ \sum_{m=2}^{k} \, \tilde{d}_{m-1} \, \tilde{A}^{(k)}_{\nu=m} \Bigr)\,.
\label{DBS.ma} 
\ee

In the case of HQ
contributions, we have a modification of (\ref{GLS.2abc}) and (\ref{GLS.2}):
\bea
&&C_{\rm{A},GLS}^{p+n}(Q^2)=
3 \,\left(1+\frac{|V_{cd}|}{1+\xi_c}-D_{\rm A}(Q^2,\xi)\right)
 +\sum_{n=2} \frac{\mu_{2n,A}}{Q^{2n-2}} \, ,
\label{Gpn.MAab.1}\\
&&\overline{C}_{\rm{A},GLS}^{p+n,(2)}(Q^2)=
3 \, \left(1+\frac{|V_{cd}|}{1+\xi_c}-D_{\rm A}(Q^2,\xi)\right)
+\frac{\hat{\mu}_{4,A} M^2}{Q^{2}+M^2} \, 
\label{Gpn.MAa.1}
\eea
where
$D_{\rm A}(Q^2,\xi)$ is obtained from 
$D(Q^2,\xi)$ in (\ref{GLS.2aa}) by replacing the derivatives $\tilde{a}^{(k)}_{\nu}$
with the corresponding analytical derivatives $\tilde{A}^{(k)}_{\nu}$.

\section{ Bjorken sum rule}

BSR \cite{Bjorken:1966jh} provides another classical observable in deep inelastic scattering, 
closely related to the GLS sum rule. Both sum rules admit an OPE
with perturbative coefficients 
and higher twist corrections, and both can therefore be treated consistently within APT.

Indeed, BSR is defined by the integral:
\be \label{BSR}
\Gamma^{p-n}_1(Q^2)=\int\limits^1_0\left[g_1^{ep}(x,Q^2)-g_1^{en}(x,Q^2)\right]dx\,.
 \ee
 Within the OPE,
 the quantity $\Gamma_1^{p-n}(Q^2)$ can be written in the form:
\bea
\Gamma_1^{p-n}(Q^2)=
\frac{g_A}{6} \, \bigl(1-D_{\rm BS}(Q^2)\bigr) + \sum_{n=2}\frac{\mu^{BS}_{2n}}{Q^{2n-2}} \, ,
\label{Gpn.OPE>BSR}
\eea
similar to (\ref{Gpn.OPEab}) for the GLS sum rule. Here,
$g_A$=1.2762 $\pm$ 0.0005 \cite{PDG20} is the axial charge of the nucleon,
$(1-D_{BS}(Q^2))$ is the leading twist contribution, and the coefficients $\mu^{BS}_{2n}$
of the higher-twist terms are free parameters that must be determined from experimental data.

Using the
"massive" twist-four representation as in (\ref{Gpn.mOPEa}), we have:
\be
\overline{\Gamma}_1^{p-n}(Q^2)=
\frac{g_A}{6} \, \bigl(1-D_{\rm BS}(Q^2)\bigr) +\frac{\hat{\mu}^{BS}_{4,A} M_{BS}^2}{Q^{2}+M_{BS}^2} \, ,
\label{Gpn.mOPE.BSR}
\ee
where the coefficients $\hat{\mu}^{BS}_4$
and $M_{BS}^2$ of the "massive"
twist-four term are free parameters that must be determined from experimental data.

Up to the $k$-th order of PT,
the twist-two parts have the form
\footnote{
  Here we consider the so-called non-singlet part in the BSR case. Note that the BSR singlet part is $\sim \sum^f_{i=1}\,e_i$ (see, e.g., \cite{Larin:2013yba}),
    where $e_i$ is the charge of the $i$ quark and $f$ is the number of active quark flavors.
    Thus, in the case $f=3$ used here, the singlet part is absent. In general, its contribution may be important
    (see \cite{Larin:2013yba,Kataev:2022iqf} and discussions therein).
}, similar to (\ref{GLS}):
\be
D^{(1)}_{\rm BS}(Q^2)=\frac{4}{\beta_0} \, a^{(1)}_s,~~D^{(k\geq2)}_{\rm BS}(Q^2)=\frac{4}{\beta_0} \, a^{(k)}_s\left(1+\sum_{m=1}^{k-1} d^{BS}_m \bigl(a^{(k)}_s\bigr)^m
\right)\,,
\label{DBS.BSR} 
\ee
where $a^{(k)}_s(Q^2)$ is defined in (\ref{ak}) and
$d^{BS}_1$,
$d^{BS}_2$,
nd $d^{BS}_3$ are exactly known (see \cite{Baikov:2010je} and references therein).

Converting the powers of the coupling constant
into its derivatives (\ref{tan+1}), we have, similarly to (\ref{GLS.1}):
\be
D^{(1)}_{\rm BS}(Q^2)=\frac{4}{\beta_0} \, \tilde{a}^{(1)}_1,~~D^{(k\geq2)}_{\rm BS}(Q^2)=
\frac{4}{\beta_0} \, \left(\tilde{a}^{(k)}_{1}+\sum_{m=2}^k\tilde{d}^{BS}_{m-1}\tilde{a}^{(k)}_{m}
\right),
\label{DBS.1.BSR} 
\ee
where the results for $\tilde{d}^{BS}_i$ $(i=1,2,3)$ can be obtained in close analogy with
(\ref{tdi}).

For the case of 3 active quark flavors ($f=3$),
we have:
\be
d^{BS}_1=\tilde{d}^{BS}_1=1.59,~~d^{BS}_2=3.99~(\tilde{d}^{BS}_2=2.73),~~d^{BS}_3=15.42~
(\tilde{d}^{BS}_3=8.61),
\ee
i.e., the coefficients in the derivative series are slightly smaller, as was the case for the GLS sum rule (see Eq. (\ref{td123})).

In APT,
the results (\ref{Gpn.OPE>BSR}) and (\ref{Gpn.mOPE.BSR}) can be written as:
\bea
&&\Gamma_{\rm{A},1}^{p-n}(Q^2)=
\frac{g_A}{6} \, \bigl(1-D_{\rm{A,BS}}(Q^2)\bigr) +\sum_{n=2} \frac{\mu^{\rm BS}_{2n,A}}{Q^{2n-2}} \, ,
\label{BSR.MAab}\\
&&\overline{\Gamma}_{\rm{A},GLS}^{p+n}(Q^2)=
\frac{g_A}{6} \, \bigl(1-D_{\rm A,BS}(Q^2)\bigr)
+\frac{\hat{\mu}^{\rm BS}_{4,A} M^2_{\rm BS}}{Q^{2}+M^2_{\rm BS}} \, ,
\label{BSR.MAa}
\eea
where the perturbative part $D_{\rm{A,BS}}(Q^2)$
takes the same forms as (\ref{DBS.1.BSR}),
but the derivatives $\tilde{a}^{(k)}_{\nu}$
  are replaced by the corresponding analytical derivatives $\tilde{A}^{(k)}_{\nu}$:
\be
D^{(1)}_{\rm A,BS}(Q^2)=\frac{4}{\beta_0} \, A^{(1)},~~
D^{k\geq2}_{\rm{A,BS}}(Q^2) =\frac{4}{\beta_0} \, \Bigl(A^{(k)}
+ \sum_{m=2}^{k} \, \tilde{d}^{\rm BS}_{m-1} \, \tilde{A}^{(k)}_{\nu=m} \Bigr)\,,
\label{DBS.ma.BSR} 
\ee

As in the GLS sum rule, the HQ
contribution for BSR was calculated in \cite{Blumlein:2016xcy},
leading to the following replacement for $\tilde{d}^{\rm BS}_1=d^{\rm BS}_1$:
\be
d^{BS}_1 \to d^{BS}_1 - \sum_{i=c,b,t}\,C_1(\xi_i),~~\tilde{d}_1 \to \tilde{d}^{BS}_1 - \sum_{i=c,b,t}\,C_1(\xi_i)\,,
    \label{Gpn.MA.HQ.BSR} 
\ee
where $\xi_i$ and $C_1(\xi)$ are given in Eqs. (\ref{xi}) and (\ref{C1}), respectively.

It is important to emphasize that the BSR perturbative structure 
is very similar to that of the GLS sum rule. Both sum rules are governed
by very similar
twist-2 parts. Their approximate equivalence has been discussed in the literature (see \cite{Broadhurst:1993ru}). Moreover, there are speculations about the similarity between the twist-four parts (see
\cite{Kataev:2005hv}).
\footnote{The similarity of perturbative parts for the structure functions $F_3(x,Q^2)$ and $g_1(x,Q^2)$ themselves has been explored in Ref. \cite{Kotikov:1996vr}.}

Therefore, in agreement with discussions in \cite{Kataev:2005hv},
we can (with some precision) treat BSR and GLS sum rule
as having the same dependence with only a different overall factor:
\be
C_{GLS}^{p+n}(Q^2)\approx \frac{18}{g_A}\,\Gamma_{1}^{p-n}(Q^2)\,
\label{GLS.app}
\ee
and, thus, for analytic QCD we have similarly:
\be
C_{\rm{A},GLS}^{p+n}(Q^2)\approx \frac{18}{g_A}\,\Gamma_{\rm{A},1}^{p-n}(Q^2)\,.
\label{GLS.app.ma}
\ee

\section{Results}

By analogy with Refs.
\cite{Pasechnik:2008th,Khandramai:2011zd,Ayala:2017uzx,Gabdrakhmanov:2023rjt,Gabdrakhmanov:2024bje,Gabdrakhmanov:2025afi}, where the BSR case was considered,
we perform a fit of experimental data for the GLS
sum rule within
the framework of standard and analytic QCD with the usual and ``massive'' forms of the twist-four terms (see Eqs. (\ref{Gpn.OPEab}), (\ref{Gpn.mOPEa}), (\ref{Gpn.MAab})
and (\ref{Gpn.MAa}), respectively).
\footnote{A similar analysis of the GLS sum rule with the usual form of the twist-four term can be found in
\cite{Milton:1998ct}.}

\begin{figure}[!htb]
\centering
\includegraphics[width=0.98\textwidth]{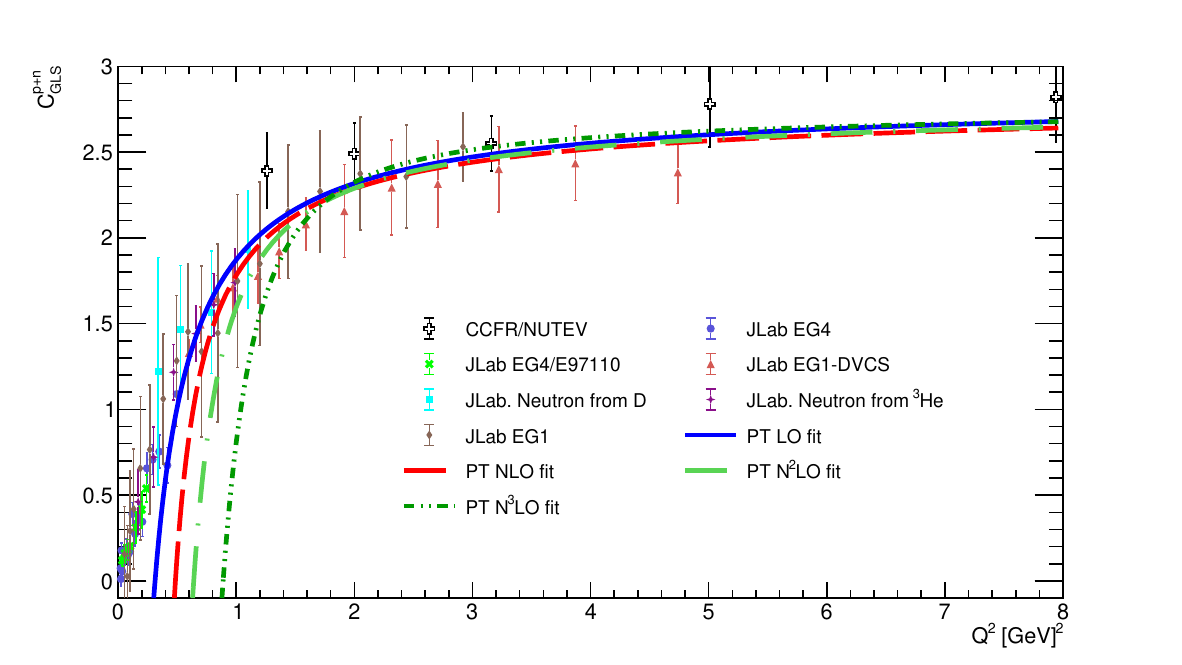}
\caption{
  \label{fig:PT1}
  The results (\ref{Gpn.OPEab}) with $\mu_{2n}=0$ for $n\geq3$
in the first three orders of PT.
}
\end{figure}

\begin{figure}[!htb]
\centering
\includegraphics[width=0.98\textwidth]{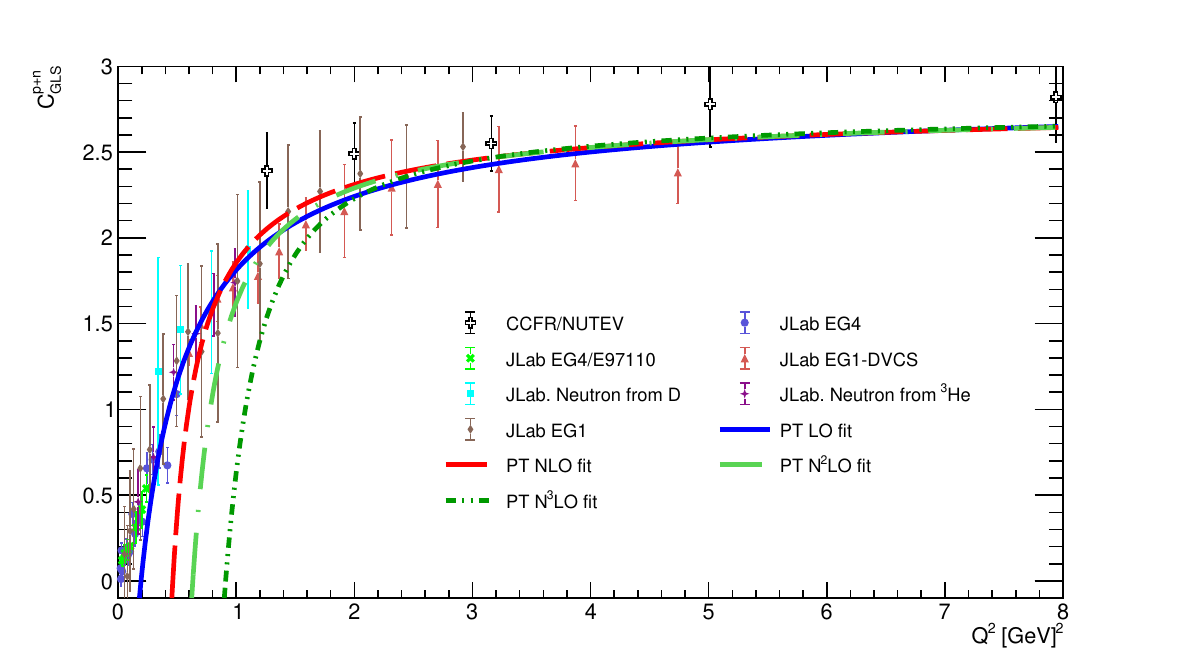}
\caption{
  \label{fig:PT2}
  As in Fig. \ref{fig:PT1} but for the massive twist-four case (\ref{Gpn.mOPEa}).
}
\end{figure}

As in the BSR case (see \cite{Gabdrakhmanov:2024bje,Gabdrakhmanov:2023rjt}), for
conventional PT, in Figs. \ref{fig:PT1} and \ref{fig:PT2}, we see a discrepancy between
the QCD predictions and the experimental data,
regardless of the form of the twist-four terms.
Moreover, we see that the discrepancy increases with increasing PT order. This is the same as in the BSR case
(see, e.g., \cite{Gabdrakhmanov:2024bje}), and the reason is the same.
With increasing PT order, the Landau pole of the strong coupling constant shifts toward higher $Q^2$ values. Thus, the results for the GLS sum rule shift toward negative values,
since the PT corrections have a negative sign.

Since the number of experimental points for
 the GLS sum rule is very small, we have added to Figs. \ref{fig:PT1} and \ref{fig:PT2} (but not used in
 the fits!)
 the experimental BSR data rescaled by a factor of $18/g_A$. 
We see that the experimental data for the GLS sum rule lie slightly above these rescaled experimental BSR points. This observation is consistent with the results of
\cite{Londergan:2010cd}, where the possibility was discussed that the experimental data for the GLS sum rule should be smaller than those published in \cite{Kim:1998kia}
due to some unconsidered corrections.

\begin{table}[!htb]
\begin{center}
\begin{tabular}{|c|c|c|c|}
\hline
& $M^2$ [GeV$^2$] & $\hat{\mu}_{\rm{MA},4}$ [ $\mu_{\rm{MA},4}$] & $\chi^2/({\rm d.o.f.})$   \\
\hline 
LO & 0.448 $\pm$ 0.041 & -2.99 $\pm$ 0.13 & 0.75  \\
&  & [-0.92 $\pm$ 0.03] & [1.63]  \\
 \hline
  NLO & 0.377 $\pm$ 0.040 & -2.95 $\pm$ 0.16 & 0.80  \\
& & [-0.80 $\pm$ 0.01] & [1.39]  \\
 \hline
N$^2$LO & 0.358 $\pm$ 0.039 & -3.00 $\pm$ 0.17 & 0.82  \\
 & & [-0.79$\pm$ 0.01] & [1.25]  \\
 \hline
 N$^3$LO & 0.354 $\pm$ 0.038 & -3.03 $\pm$ 0.17 & 0.82  \\
  & & [-0.78 $\pm$ 0.03] & [1.25]  \\
 \hline
\end{tabular}
\end{center}
\caption{%
  The values of the fit parameters for the massive twist-four case shown in (\ref{Gpn.MAa})
  [for the massless one shown in (\ref{Gpn.MAab})].
}
\label{Tab:GLSlp}
\end{table}

\begin{figure}[!htb]
\centering
\includegraphics[width=0.98\textwidth]{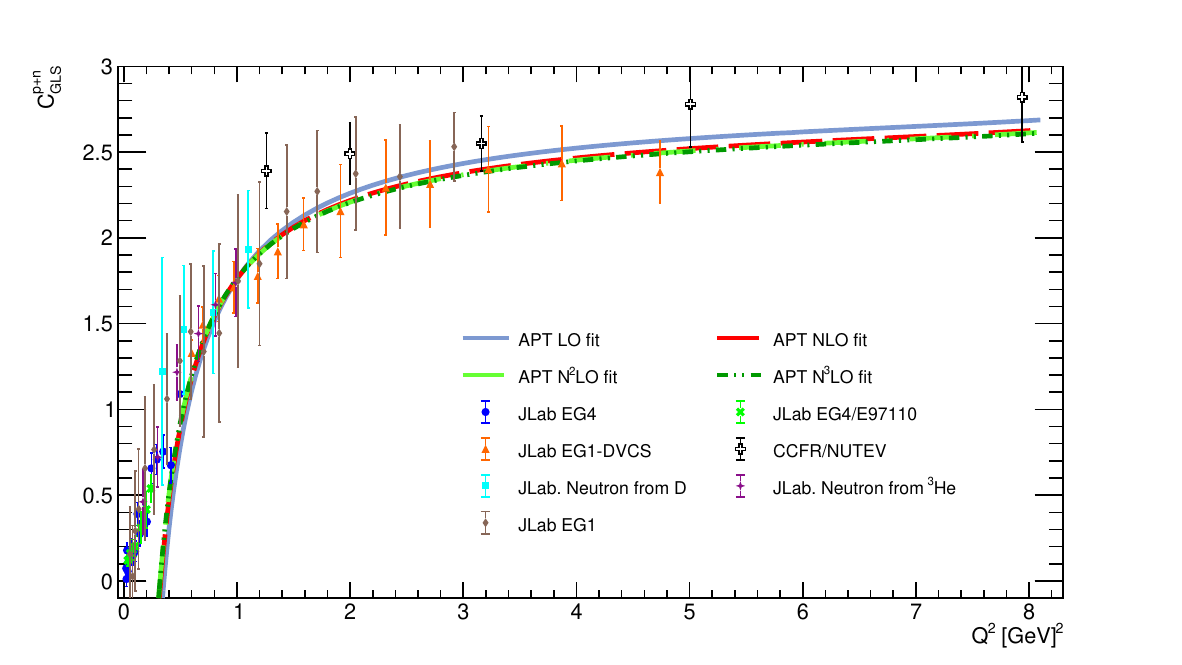}
\caption{
  \label{fig:APT2}
  As in Fig. \ref{fig:PT1} but for the APT result (\ref{Gpn.MAab})
}
\end{figure}

\begin{figure}[!htb]
\centering
\includegraphics[width=0.98\textwidth]{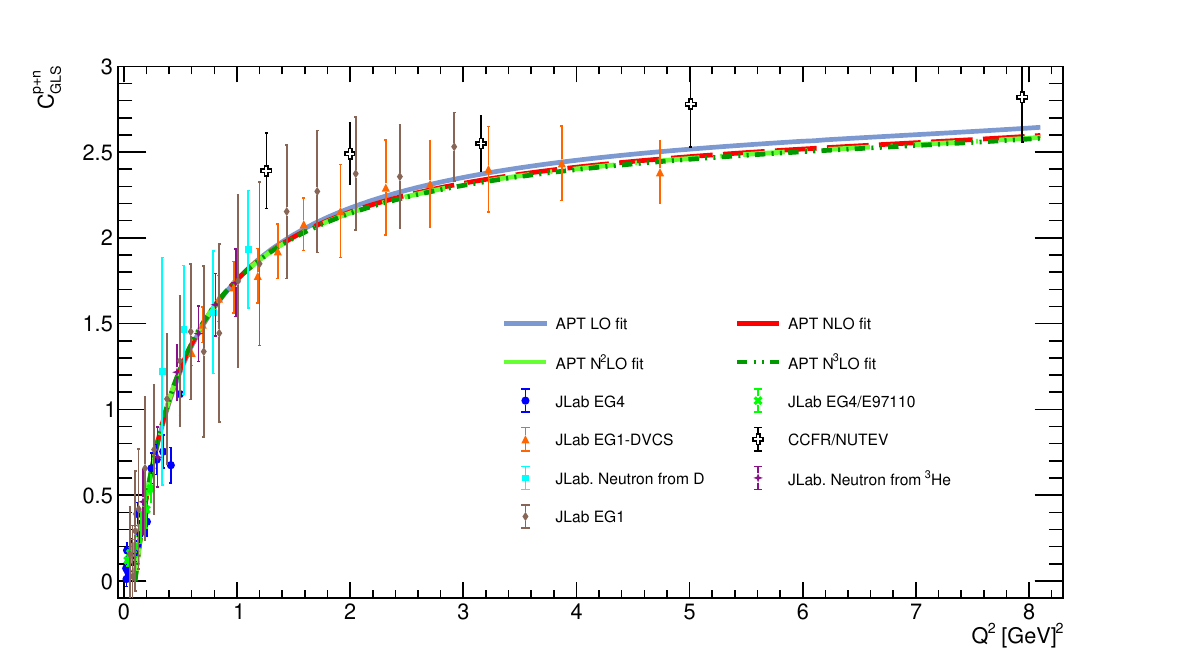}
\caption{
  \label{fig:APT1}
  As in Fig. \ref{fig:PT2} but for the APT result (\ref{Gpn.MAa})
}
\end{figure}

In the case of analytic QCD, we have good agreement between the QCD predictions and the experimental data
(see Table \ref{Tab:GLSlp}
and Figs. \ref{fig:APT2} and \ref{fig:APT1}).
In the figures
we also show the experimental BSR data rescaled by the factor $18/g_A$.
With the massive form of the twist-4 term (see Fig. \ref{fig:APT1}), the agreement with the
rescaled BSR data
extends down to the low-$Q^2$ region, remaining good for $Q^2<0.6$ GeV$^2$. In contrast, the "standard" twist-4 ansatz
deteriorates at low $Q^2$ (see Fig. \ref{fig:APT2}) and provides a reasonable description only
for $Q^2 \geq 0.8$ GeV$^2$.
Of course, this correspondence occurs only for the rescaled BSR data, but
not for the data according to the GLS sum rule, which are absent at such low values of $Q^2$.

The available experimental information on the GLS sum rule still has large uncertainties,
and the number of experimental points is very small.
For this reason, it is useful to
supplement the comparison with recent lattice QCD results from Ref. \cite{Can:2025zsr}.
Note that the lattice data in \cite{Can:2025zsr} are quite different for different values of $\beta$. We will use the lattice data with a large $\beta$ corresponding to the smallest value of
the lattice parameter $a$.

\begin{figure}[!htb]
\centering
\includegraphics[width=0.98\textwidth]{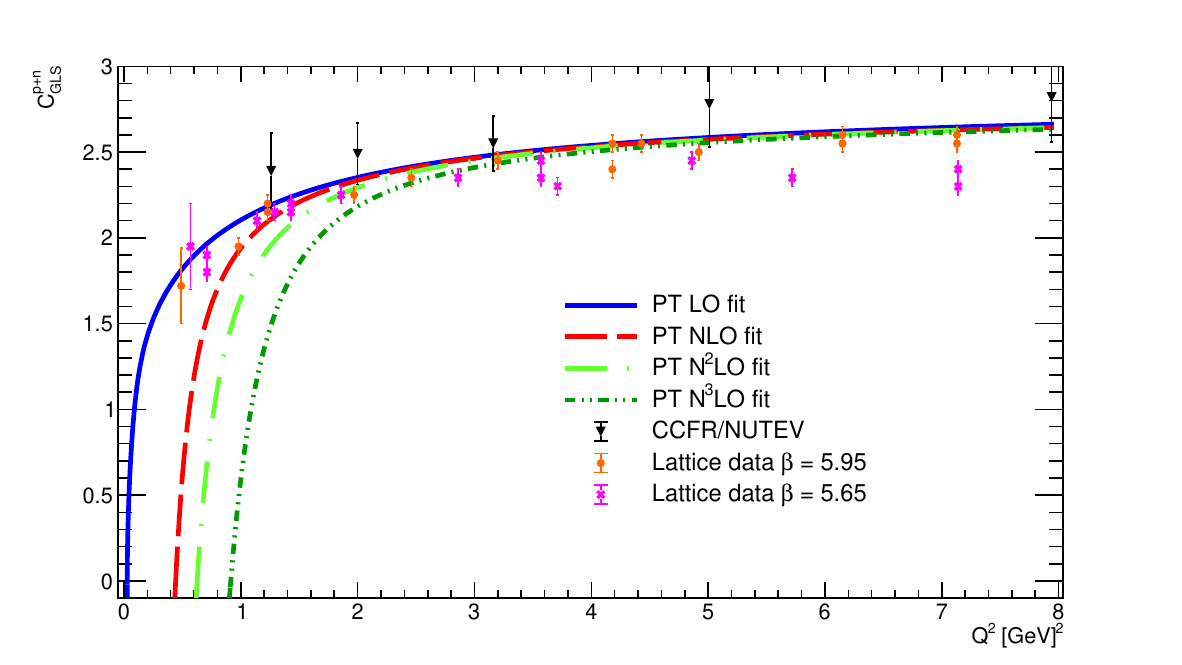}
\caption{
  \label{fig:PT1L}
  As in Fig. \ref{fig:PT2} but with lattice data.
}
\end{figure}

In the case of standard PT, the results obtained by fitting these lattice ``data'' are presented in
Fig. \ref{fig:PT1L}. (Since the results in the cases of the usual and massive twist-four terms are similar,
we show only the results for the massive twist-four term).
It can be seen that the LO results agree quite well with the lattice "data", which cannot be said about the results for higher PT orders.
\footnote{
Given the close analogy with the BSR case (see \cite{Gabdrakhmanov:2024vol}), a detailed analysis of the low $Q^2$ range
can lead to an incorrect description, even at LO. Unfortunately, in the case of the GLS sum rule,
there is no information about the low $Q^2$ range, since the relevant experimental and lattice
``data'' are missing here.
}

Moreover, as already discussed for the case of the experimental data for the GLS sum rule, the
discrepancy increases with increasing PT order. The reasons are the same as discussed above.

\begin{table}[!htb]
\begin{center}
\begin{tabular}{|c|c|c|c|}
\hline
& $M^2$ [GeV$^2$]  & $\hat{\mu}_{\rm{A},4}$  & $\chi^2/({\rm d.o.f.})$   \\
& ($+$ charm contributions) & ($+$ charm contributions) &  ($+$ charm contributions)\\
& [$+$ HQ contributions] &  [$+$ HQ contributions] &  [$+$ HQ contributions] \\
\hline 
LO & 0.879 $\pm$ 0.338 & -1.37 $\pm$ 0.31 & 4.84  \\
& (1.493 $\pm$ 0.391) & (-2.10 $\pm$ 0.61) & (4.51)  \\
  & [1.253 $\pm$ 0.432] & [-1.88 $\pm$ 0.61] & [4.79]  \\
 \hline
  NLO & 0.408 $\pm$ 0.277 & -1.54 $\pm$ 0.32 & 3.50  \\
& (0.886 $\pm$ 0.400) & (-2.01 $\pm$ 0.46) & (4.48)  \\
  & [0.998 $\pm$ 0.337] & [-1.83 $\pm$ 0.51] & [4.15]  \\
 \hline
N$^2$LO & 0.400 $\pm$ 0.258 & -1.68$\pm$ 1.06 & 3.24  \\
 & (0.785 $\pm$ 0.385) & (-2.11$\pm$ 0.46) & (4.04)  \\
& [0.878 $\pm$ 0.319] & [-1.93$\pm$ 0.59] & [4.48]  \\
 \hline
 N$^3$LO & 0.370 $\pm$ 0.247 & -1.78 $\pm$ 1.20 & 3.12  \\
  & (0.738 $\pm$ 0.373) & (-2.20 $\pm$ 0.77) & (4.34)  \\
 & [0.822 $\pm$ 0.306] & [-2.02 $\pm$ 0.51] & [4.02]  \\
   \hline
 \hline
\end{tabular}
\end{center}
\caption{%
  The values of the parameters in (\ref{Gpn.MAa}) and (\ref{Gpn.MAa.1})
obtained from the lattice data.
}
\label{Tab:GLSla1}
\end{table}

\begin{figure}[!htb]
\centering
\includegraphics[width=0.98\textwidth]{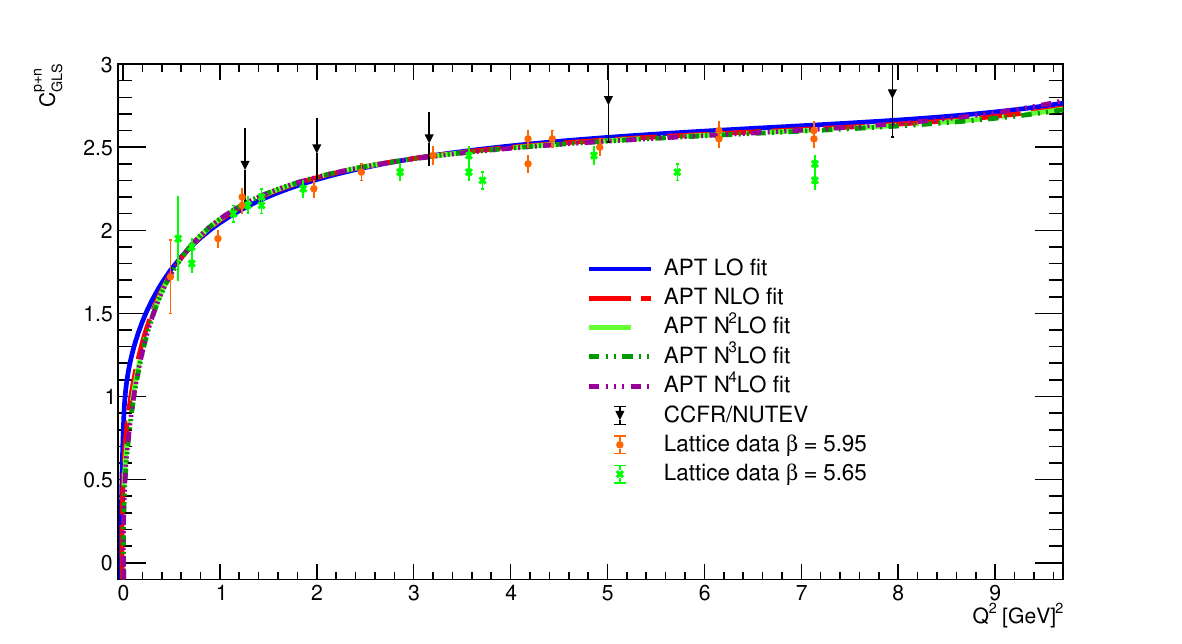}
\caption{
  \label{fig:APT2L}
  As in Fig. \ref{fig:APT1} but with the lattice data.
}
\end{figure}

In the case of analytic QCD, we have good agreement with the lattice ``data''. 
The results obtained by fitting these lattice data are presented in Table \ref{Tab:GLSla1} and
Fig. \ref{fig:APT2L}.
(The results in the cases of the usual and massive twist-four terms are similar; therefore,
we show only the results for the massive twist-four term.)
The fitting parameters obtained with massless
light quarks, with the addition
of the charm contribution, and with the addition of the contribution of all heavy quarks are quite similar. 
Moreover, since there is no real difference between the results for the massless and massive
twist-four terms, we present in Table \ref{Tab:GLSla1} only the results for the massive twist-four term.

\begin{table}[!htb]
\begin{center}
\begin{tabular}{|c|c|c|c|}
\hline
& $M^2$ [GeV$^2$] for the massless case & $\hat{\mu}_{\rm{A},4}$  for the massless case & $\chi^2/({\rm d.o.f.})$  the massless case  \\
& ($+$ charm contributions) & ($+$ charm contributions) &  ($+$ charm contributions)\\
& [$+$ HQ contributions] &  [$+$ HQ contributions] &  [$+$ HQ contributions] \\
\hline 
LO & 1.283 $\pm$ 0.819 & -1.10 $\pm$ 0.31 & 3.18  \\
& (1.466 $\pm$ 0.592) & (-1.74 $\pm$ 0.33) & (3.16)  \\
  & [1.699 $\pm$ 0.595] & [-1.70 $\pm$ 0.29] & [3.34]  \\
 \hline
  NLO & 0.609 $\pm$ 0.556 & -1.32 $\pm$ 0.25 & 2.58  \\
& (1.108 $\pm$ 0.463) & (-1.74 $\pm$ 0.42) & (2.75)  \\
  & [1.225 $\pm$ 0.512] & [-1.69 $\pm$ 0.38] & [2.94]  \\
 \hline
N$^2$LO & 0.477 $\pm$ 0.491 & -1.45$\pm$ 1.06 & 2.41  \\
 & (1.006 $\pm$ 0.435) & (-1.79$\pm$ 0.45) & (2.43)  \\
& [1.143 $\pm$ 0.481] & [-1.72$\pm$ 0.41] & [2.84]  \\
 \hline
 N$^3$LO & 0.429 $\pm$ 0.457 & -1.57 $\pm$ 1.20 & 2.33  \\
  & (0.957 $\pm$ 0.415) & (-1.85 $\pm$ 0.47) & (2.59)  \\
 & [1.088 $\pm$ 0.458] & [-1.76 $\pm$ 0.43] & [2.79]  \\
   \hline
 \hline
\end{tabular}
\end{center}
\caption{%
  The values of the parameters in (\ref{Gpn.MAa}) and (\ref{Gpn.MAa.1})
obtained from experimental and lattice data.
}
\label{Tab:GLSla+}
\end{table}

Note that we obtain rather large $\chi^2$ values for the fits because the uncertainties in the lattice ``data'' are very small.
If we add experimental data to the considered `data' of the lattice, the values of $\chi^2$ decrease,
but the extracted parameters do not actually change when taking into account the corresponding uncertainties
(see Table \ref{Tab:GLSla+}). 

\begin{figure}[t]
\centering
\includegraphics[width=0.98\textwidth]{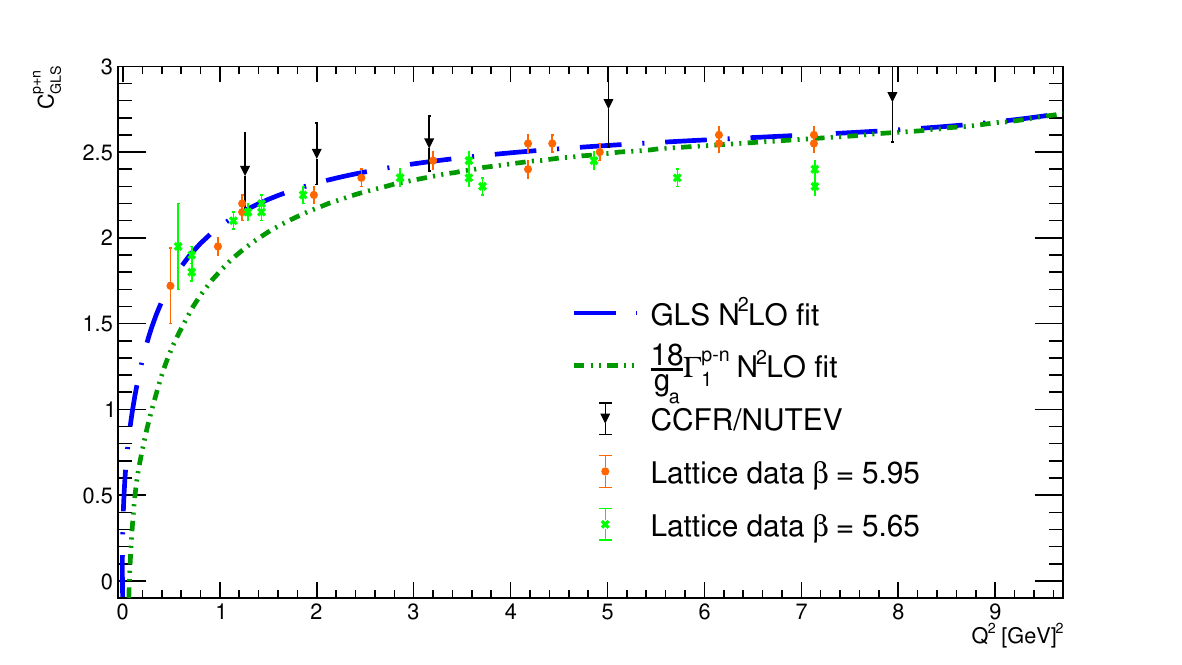}
\caption{
  \label{fig:low1}
  As in Fig. \ref{fig:APT2L} with the rescaled BSR curve (\ref{GLS.app.ma}) added.
}
\end{figure}

In Fig. \ref{fig:low1}, we show the results for the GLS sum rule obtained in Table
\ref{Tab:GLSla1}, along with the BSR results scaled by a factor of $18/g_A$.
The BSR results were obtained under the same conditions in Ref. \cite{Gabdrakhmanov:2024bje}. 
As can be seen from Fig. \ref{fig:low1}, the ratio (\ref{GLS.app.ma}) is valid only for large
values of $Q^2$: $Q^2\geq$5 GeV$^2$. For smaller values of $Q^2$, $C_{\rm{A},GLS}^{p+n}(Q^2)$ and
$(18/g_A)\,\Gamma_{\rm{A},1}^{p-n}(Q^2)$ become different. The difference arises due to the different parameter values
in the twist-four terms in the cases of the GLS and Bjorken sum rules. 
Indeed, the values of $\hat{\mu}_{\rm{A},4}$ shown in Table \ref{Tab:GLSla1} are significantly smaller than
the corresponding values for the twist-four magnitudes obtained for the BSR case (see Table I in
\cite{Gabdrakhmanov:2024bje}) and rescaled by a factor of $(18/g_A)$.
Thus, the ratio 
(\ref{GLS.app.ma}), based on the assumption in \cite{Kataev:2005hv}, seems to be fulfilled only
for rather large values of $Q^2$, $Q^2\geq$ 5 GeV$^2$.

\section{Conclusions}

We have investigated the Gross-Llewellyn Smith sum rule within the framework of conventional and analytic QCD.
A comparison was performed between experimental data, recent lattice calculations, and perturbative QCD
based on conventional and analytic versions of perturbation theory with different
parametrizations for the twist-four contribution. We showed that conventional perturbation
theory 
fails to reproduce the data, while
the analytic version, especially with the massive twist-four term, demonstrates good agreement with experimental data
and lattice calculations.

We also discussed the relation between the Gross-Llewellyn Smith sum rule and the Bjorken sum rule. We found a similarity
between the Gross-Llewellyn Smith sum rule and the rescaled Bjorken sum rule, with the additional factor $18/g_A$,
at rather large $Q^2$ values, $Q^2 \geq 5$ GeV$^2$, where the corresponding twist-two parts give the main contributions. 
However, at smaller $Q^2$ values ($Q^2 \leq 2$ GeV$^2$), there is a difference between them.
We have shown that this is due to the different values of the twist-four terms obtained here for the GLS sum
rule (see Table \ref{Tab:GLSla1}) and for the BSR case in Ref. \cite{Gabdrakhmanov:2024bje}.


{\bf Acknowledgments.}~The authors are grateful to
Johannes Blumlein for initiating the consideration of the contribution of heavy quarks.
One of us (I.A.Z.)
is supported by ANID grants Fondecyt Regular N1251975.

\end{document}